\documentstyle[aps,preprint,12pt]{revtex}

\def\mathbb#1{{\cal #1}\kern-0.6em I \kern 0.2em}
\draft
\title{QUANTUM EINSTEIN'S EQUATIONS AND CONSTRAINTS ALGEBRA}
\author{FATIMAH SHOJAI\thanks{Email: FATIMAH@THEORY.IPM.AC.IR}}
\address{Physics Department, Iran University of Science and Technology, P.O.Box 16765--163, Narmak, Tehran,
IRAN}
\address{and}
\address{Institute for Studies in Theoretical Physics and Mathematics, P.O.Box 19395-5531, Tehran,
IRAN}
\author{ALI SHOJAI\thanks{Email: SHOJAI@THEORY.IPM.AC.IR}}
\address{Physics Department, Tarbiat Modares University,P.O.Box 14155--4838, Tehran, IRAN}
\address{and}
\address{Institute for Studies in Theoretical Physics and Mathematics, P.O.Box 19395-5531, Tehran,
IRAN}
\begin{document}
\maketitle
\begin{abstract}
In this paper we shall address this problem: Is quantum gravity
constraints algebra closed and what are the quantum Einstein
equations. We shall investigate this problem in the
de-Broglie--Bohm quantum theory framework. It is shown that the
constraint algebra is weakly closed and the quantum Einstein's
equations are derived.
\end{abstract}
\section{Introduction}
de-Broglie--Bohm causal quantum mechanics\cite{BOHM} has several
positive points. (a) First of all, it is causal and so describes the system in an ordered way in time. (b) Perhaps the most important concept in the
de-Broglie--Bohm theory is the lack of need for the assumption of
the existence of a classical domain in the measurement phenomena.
(c) This theory provides a useful framework for quantum gravity. (d) It does not suffer from the conceptual problems like the meaning of the wavefunction for a single system and so on. In fact there is a
number of such positive points of this theory which can be found
in the literature\cite{POS}.

The application of this theory to quantum gravity has been
investigated from different aspects: The quantum force may be
repulsive and thus can remove the initial singularity and
inflation would be emerged; Since the quantum domain is defined
as the domain that the quantum force is smaller than the classical
force, it is possible to have the classical universe for the
small scale factors and conversely quantum universe for large
scale factors\cite{PIN1}. As the scale factor represents the universe radius, there is not a one--to--one correspondence between large and classical universes; Because of the guidance formula the
time parameter appears automatically and the time problem does not
exist; Moreover a new approach based on the de-Broglie--Bohm
theory has been presented that brings many interesting physical
results, such as unification of quantal and gravitational
behaviour of matter\cite{FATI1}.

In the ADM decomposition of the space--time in general relativity
the non--dynamical nature of shift and lapse functions (these are
functions used in the slicing of the space--time and will be
introduced later) must be consistent with the evolution. This is
obviously satisfied if the constraint algebra be closed. Moreover
as we know the secondary constraint in the ADM decomposition of
the space--time are energy and momentum constraints. These are
four non dynamical Einstein's equations which together with the
closedness of constraint algebra provides necessary and
sufficient conditions for lapse and shift function to be
non--dynamical as the space--time evolves. This is satisfied at
the classical level as we know that the general relativity is
independent of the space--time reparametrization. This property
is guaranteed by the geometrical Bianchi identities.

In the present work we want to discuss this point at the quantum
level. In order to do so, first we shall derive the constraint
algebra at the quantum level. We shall use the integrated version
of diffeomorphism and hamiltonian constraints. Next in the other
part we shall derive the quantum Einstein's equations choosing
arbitrary lapse and shift functions i.e. in an arbitrary gauge.
These equations are the extended form of those previously
discussed by us\cite{Covariance}.

\section{Quantum Constraints Algebra}
In the ADM formulation, the hamiltonian of general relativity is:
\begin{equation}
H=\int d^3x {\cal H}
\end{equation}
in which
\begin{equation}
{\cal H}=N{\cal H}_0+N^i{\cal H}_i
\end{equation}
where
\begin{equation}
{\cal H}_0=G_{ijkl}\Pi^{ij}\Pi^{kl}+\sqrt{h}({\cal R}-2\Lambda)
\end{equation}
and
\begin{equation}
{\cal H}_i=-2\nabla_j\Pi_i^j
\end{equation}
In these equations, $\vec{N}$ is the shift function and $N$ is the
lapse function. These functions produce the time evolution of
space--like surfaces in normal and tangent directions
respectively. $G_{ijkl}$ is the superspace metric and is given by:
\begin{equation}
G_{ijkl}=\frac{1}{2\sqrt{h}}\left (
h_{ik}h_{jl}+h_{il}h_{jk}-h_{ij}h_{kl}\right )
\end{equation}
$h_{ij}$ is the three dimensional space--like hypersurface metric
and $\Pi^{ij}$ is the conjugate canonical momentum of $h_{ij}$,
${\cal R}$ is the intrinsic curvature of the hypersurface, $\Lambda$
is the cosmological constant. By Guass--Codazzi relations one can
show that 
\begin{equation}
{\cal H}_0=0
\label{zxw1}
\end{equation}
and 
\begin{equation}
{\cal H}_i=0
\label{zxw2}
\end{equation}
 are in fact the four
non--dynamical (constraints) Einstein equations. So the
hamiltonian density vanishes by the Einstein's equations. The other dynamical Einstein's equations can be derived by differentiating the first equation with respect to $h_{ij}$.

Corresponding to the two directions of time evolution of
hypersurface (normal and tangent), the hamiltonian can be
separated as:
\begin{equation}
\mathbb{C}_c(N)=\int d^3x N{\cal H}_0
\end{equation}
and
\begin{equation}
\widetilde{\mathbb{C}}_c(\vec{N})=\int d^3x N^i{\cal H}_i
\end{equation}
Because of the above explanations,
$\widetilde{\mathbb{C}}_c(\vec{N})$ is called diffeomorphism
constraint while $\mathbb{C}_c(N)$ is called hamiltonian
constraint. It is instructive to see the algebra of constraints.
Using the notation of ref.\cite{Baez}, one can obtains:
\begin{equation}
\left
\{\widetilde{\mathbb{C}}_c(\vec{N}),\widetilde{\mathbb{C}}_c(\vec{N}')\right
\}=\widetilde{\mathbb{C}}_c(N^i\vec{\nabla}N'_i-N'^i\vec{\nabla}N_i)
\label{a}
\end{equation}
\begin{equation}
\left \{\mathbb{C}_c(N),\mathbb{C}_c(N')\right
\}=\widetilde{\mathbb{C}}_c(N\vec{\nabla}N'-N'\vec{\nabla}N)
\end{equation}
\begin{equation}
\left \{\widetilde{\mathbb{C}}_c(\vec{N}),\mathbb{C}_c(N)\right
\}=\mathbb{C}_c(\vec{N}\cdot\vec{\nabla}N)
\end{equation}
To quantize according to the standard quantum mechanics, one can
use the Dirac quantization procedure which leads to:
\begin{equation}
\widehat{\mathbb{C}(N)}\Psi=0
\end{equation}
\begin{equation}
\widehat{\widetilde{\mathbb{C}}(\vec{N})}\Psi=0
\end{equation}
These are quantum constraint and limit the physical wave function.
The former is WDW equation and the latter represents the
invariance under general spatial transformation.

Now we shall apply the de-Broglie--Bohm theory to canonical
quantum gravity. In the Hamilton--Jacobi language (which is
suitable for our discussion), in de-Broglie--Bohm theory the
desired quantum system is subjected to quantum potential in
addition to the classical ones. This term includes all the quantum
information about the system. It is a non--local potential and
obtained by the norm of the wave--function. By this simple
description of this theory, we can discuss the constraints algebra
at the quantum level.

As discussed in refs.\cite{POS,Covariance}, the following change in ${\cal H}_0$ alone will be sufficient for description at quantum level:
\begin{equation}
{\cal H}_0\rightarrow {\cal H}_0+Q
\end{equation}
where $Q$ is the quantum potential. 
The constraint equations corresponding to equations (\ref{zxw1}) and (\ref{zxw2}) are:
\begin{equation}
G_{ijkl}\Pi^{ij}\Pi^{kl}+\sqrt{h}({\cal R}-2\Lambda)+Q=0
\label{qwe1}
\end{equation}
\begin{equation}
-2\nabla_j\Pi_i^j=0
\label{qwe2}
\end{equation}
Dynamical equations are derived by differentiating the first equation and using Gauss--Codazzi relations, as this is done in the next section.

Equivalently we have:
\begin{equation}
\mathbb{C}(N)=\mathbb{C}_c(N)+{\cal Q}(N)
\end{equation}
\begin{equation}
\widetilde{\mathbb{C}}(\vec{N})=\widetilde{\mathbb{C}}_c(\vec{N})
\end{equation}
where
\begin{equation}
{\cal Q}(N)=\int d^3x N Q=\int d^3x N \left (
-\frac{\hbar^2\kappa}{\sqrt{h}|\Psi|}G_{ijkl}\frac{\delta^2|\Psi|}{\delta
h_{ij}\delta h_{kl}} \right )
\end{equation}
Since at the quantum level $\widetilde{\mathbb{C}}(\vec{N})$ has
not any change the relation (\ref{a}) is satisfied again. Some
calculations leads to the following algebra relations:
\begin{equation}
\left
\{\widetilde{\mathbb{C}}(\vec{N}),\widetilde{\mathbb{C}}(\vec{N}')\right
\}=\widetilde{\mathbb{C}}(N^i\vec{\nabla}N'_i-N'^i\vec{\nabla}N_i)
\label{b}
\end{equation}
\[\left \{\mathbb{C}(N),\mathbb{C}(N')\right
\}=\widetilde{\mathbb{C}}_c(N\vec{\nabla}N'-N'\vec{\nabla}N)+\]
\begin{equation}
2\int d^3z d^3x \sqrt{h(z)}G_{ijkl}(z)\Pi^{kl}(z)\left(
-N(z)N'(x)+N(x)N'(z)\right )\frac{\delta Q(x)}{\delta
h_{ij}(z)}\approx 0 \label{c}
\end{equation}
\begin{equation}
\left \{\widetilde{\mathbb{C}}(\vec{N}),\mathbb{C}(N)\right
\}=\mathbb{C}(\vec{N}\cdot\vec{\nabla}N) \label{d}
\end{equation}
The relation (\ref{d}) is the same as the classical one and the relation
(\ref{c}) is weakly zero ($\approx 0$), i.e. zero only when the
equation of motion is used. 
To see this, one must evaluate the derivative of quantum potential from the equation of motion (\ref{qwe1}) as:
\begin{equation}
\frac{\delta Q(x)}{\delta h_{ij}(z)}=\frac{3}{4\sqrt{h}}h_{kl}\Pi^{ij} \Pi^{kl} \delta (x-z) -\frac{\sqrt{h}}{2}h^{ij}({\cal R}-2\Lambda)\delta (x-z) -\sqrt{h} \frac{\delta {\cal R}}{\delta h_{ij}}
\end{equation}
Using the known identity:
\begin{equation}
F\frac{\delta {\cal R}(x)}{\delta h_{ij}(z)}=\left ( -F{\cal R}^{ij}+ \nabla^i \nabla^j F-h^{ij}\nabla^2F\right )\delta (x-z)
\end{equation}
where $F$ is any arbitrary function, and
substituting this relations in the poisson bracket (\ref{c}), shows that it is weakly zero.
This means the one parameter family of
diffeomorphism of spatial slices is a symmetry of the quantum
space--time. But pushing spatial slices in normal direction is a
symmetry satisfied only at the level of quantum equations of
motion. This point confirms our previous result
(\cite{Covariance} and its references).

Therefore as at the classical level, we must choose the initial
conditions for space-like metric, lapse and shift functions such
that to be consistent with the symmetry of space-time. Since the
constraint algebra isn't closed strongly, the hamiltonian isn't
invariant under reparametrization of space-time. Then under
dynamical evolution the symmetry dosen't survive. In this way
different identical conditions leads to different solutions.

Because of the lack of invariance under time reparametrization we
expect that the general covariance symmetry be broken, but this
is not obvious at the level of the equations of motion. If we are
interested in seeing the symmetry breaking at this level, we must
look at quantum Einstein's equations. This point is the content
of the next section.
\section{Quantum Einstein Equations}
Previously\cite{Covariance} in the Bohmian quantum gravity
framework, we have studied the modifications of Einstein's
equations in some special gauge. It was shown that the correction
terms contain the quantum potential as we expected. Our discussion
in\cite{Covariance} is based on ADM decomposition of the
space--time and Gauss-Codazzi equations. But it was assumed there,
the lapse function is 1 and the shift is zero, for
simplification. Here we derive the modified Einstein's equations
in the general case.

The Gauss-Codazzi equations for any choice of lapse and shift
functions are
\begin{equation}
G_{\mu\nu}n^\mu n^\nu=-\frac{1}{2}\left ( {\cal
R}+K^2-K_{ij}K^{ij}\right ) \label{f}
\end{equation}
and
\begin{equation}
G_{\mu i}n^\mu = \nabla_jK^j_i- \nabla_iK^j_i
\end{equation}
where
\begin{equation}
n_\mu=\frac{1}{N}(1,-\vec{N})
\end{equation}
represents a field of time--like vectors normal to a space--like
slice. $K_{ij}$ and ${\cal R}_{ij}$ are extrinsic curvature and
Riemann tensor of three metric respectively. If we express the
constraints quantities, $\mathbb{C}$ and $\widetilde{\mathbb{C}}$,
in terms of the intrinsic curvature, using the following formula
for conjugate momenta:
\begin{equation}
\Pi^{ij}=\sqrt{h}(K^{ij}-h^{ij}K)
\end{equation}
we find:
\begin{equation}
{\cal H}_0=-2G_{\mu\nu}n^\mu n^\nu
\end{equation}
\begin{equation}
{\cal H}_i=-2G_{\mu i}n^\mu
\end{equation}
Now according to de-Broglie--Bohm quantum theory of gravity we
have:
\begin{equation}
-2G_{\mu\nu}n^\mu n^\nu-Q=0 \label{h}
\end{equation}
\begin{equation}
G_{\mu i}n^\mu=0 \label{i}
\end{equation}
In continuation we use the form of general relativity action in
terms of three metric and extrinsic curvature:
\begin{equation}
{\cal A}=-\frac{1}{16\pi G}\int d^4x \sqrt{-g}( {\cal
R}+K^2-K_{ij}K^{ij}) \label{e}
\end{equation}
Using the relations (\ref{e}), (\ref{f}), and (\ref{h}) the
dynamical equation of three metric is obtained:
\begin{equation}
{\cal G}^{ij}=\frac{\delta {\cal Q}}{\delta h_{ij}}
\end{equation}
As we expected in the right hand side of the above dynamical
equation the quantum force is appeared. From this relation by
using equation (\ref{h}) and (\ref{i}) we get the other remaining
Einstein equations:
\begin{equation}
{\cal G}^{0i}=NN^iQ+N_j\frac{\delta {\cal Q}}{\delta h_{ij}}
\end{equation}
\begin{equation}
{\cal G}^{00}=N\left (N_iN^i-\frac{N}{2}\right
)Q+N_iN_j\frac{\delta {\cal Q}}{\delta h_{ij}}
\end{equation}
It is simply seen that these equations are general form of the
results of ref\cite{Covariance}. Thus in the modified Einstein's
equations in a general case both shift and lapse functions appear.
This motivates us to doubt about covariance under general
coordinate transformations. A simple calculation shows that these
equations aren't covariant. 
Setting the right hand side of modified Einstein's equations as the components of a matrix ${\cal X}^{\mu\nu}$, to show the lack of general covariance one must investigate the transformation properties of ${\cal X}^{\mu\nu}$. Specifying the transformation to one in which $t\rightarrow t'(t)$ and $\vec{x}$ does not change, we have (using the transformation law of the metric):
\begin{equation}
h'_{ij}=h_{ij}
\end{equation}
\begin{equation}
N'=FN
\end{equation}
\begin{equation}
N'^i=FN^i
\end{equation}
with:
\begin{equation}
F=\frac{\partial t}{\partial t'}
\end{equation}
and the quantum potential would remain unchanged. So we have:
\begin{equation}
{\cal X}'^{00}=FN(F^2N_iN^i-\frac{FN}{2})Q+F^2N_iN_j\frac{\delta {\cal Q}}{\delta h_{ij}}
\end{equation}
\begin{equation}
{\cal X}'^{0i}=F^2NN^iQ+FN_j\frac{\delta {\cal Q}}{\delta h_{ij}}
\end{equation}
\begin{equation}
{\cal X}'^{ij}={\cal X}^{ij}
\end{equation}
which shows that ${\cal X}^{\mu\nu}$ does not transform as a second rank tensor.
Thus the general covariance principle
is broken at the quantum level.
\section{conclusion}
The quantum effects can be studied in gravity, as well as any
other theory, by introducing the quantum potential. Since the
quantum potential modifies the hamiltonian, there is no guarantee
that the constraints algebra be closed as it is in the classical
case. The constraints algebra is in fact closed only weakly, i.e.
by using the equations of motion. This shows that the associated
symmetry should break down. We saw that how one can obtain the
equations of motion, i.e. the quantum Einstein's equations and
how they are not general covariant.

\end{document}